\documentclass[conference]{IEEEtran}
\IEEEoverridecommandlockouts
\usepackage{amsmath,graphicx}
\usepackage{amsmath,amssymb,amsfonts}
\usepackage{cite, soul}
\usepackage{url}
\usepackage{amsthm}
\usepackage{bm}
\usepackage{comment}
\usepackage{multirow}
\usepackage{booktabs}
\usepackage{hhline}
\usepackage{graphicx}     
\usepackage{caption}      
\usepackage{subcaption}   
\usepackage{setspace}

\usepackage[symbol]{footmisc}
\renewcommand{\thefootnote}{\fnsymbol{footnote}}
\newcommand \footnoteONLYtext[1]
{
	\let \mybackup \thefootnote
	\let \thefootnote \relax
	\footnotetext{#1}
	\let \thefootnote \mybackup
	\let \mybackup \imareallyundefinedcommand
}

\usepackage{setspace}
\usepackage{setspace}
\usepackage[ruled,linesnumbered]{algorithm2e}

\DeclareMathOperator*{\argmax}{argmax}
\usepackage[usenames,dvipsnames]{color}

\usepackage[top=0.7in, bottom=0.98in, left=0.627in, right=0.620in]{geometry}
\def\@#1{\pmb{#1}}
\def\b#1{\mathbb{#1}}
\def\bf#1{\mathbf{#1}}
\def\s#1{\mathsf{#1}}

\title{Multimodal Deep Learning-Empowered Beam Prediction in Future THz ISAC Systems}
\author{Kai Zhang, Wentao Yu, Hengtao He, \emph{Member}, \emph{IEEE}, Shenghui Song, \emph{Senior Member}, \emph{IEEE},\\ Jun Zhang, \emph{Fellow}, \emph{IEEE}, and Khaled B. Letaief, \emph{Fellow}, \emph{IEEE}\\
Dept. of ECE, The Hong Kong University of Science and Technology, Hong Kong\\
Email: \{kzhangbn, wyuaq\}@connect.ust.hk, \{eehthe, eeshsong, eejzhang, eekhaled\}@ust.hk
}

\begin{document}


\maketitle

\footnoteONLYtext{
	\hspace*{-1.599em}
	This work is supported in part by the Hong Kong Research Grant Council under Grant No. 16209023.
}

\begin{abstract}

Integrated sensing and communication (ISAC) systems operating at terahertz (THz) bands are envisioned to enable both ultra-high data-rate communication and precise environmental awareness for next-generation wireless networks. However, the narrow width of THz beams makes them prone to misalignment and necessitates frequent beam prediction in dynamic environments. Multimodal sensing, which integrates complementary modalities such as camera images, positional data, and radar measurements, has recently emerged as a promising solution for proactive beam prediction. Nevertheless, existing multimodal approaches typically employ static fusion architectures that cannot adjust to varying modality reliability and contributions, thereby degrading predictive performance and robustness. To address this challenge, we propose a novel and efficient multimodal mixture-of-experts (MoE) deep learning framework for proactive beam prediction in THz ISAC systems. The proposed multimodal MoE framework employs multiple modality-specific expert networks to extract representative features from individual sensing modalities, and dynamically fuses them using adaptive weights generated by a gating network according to the instantaneous reliability of each modality. Simulation results in realistic vehicle-to-infrastructure (V2I) scenarios demonstrate that the proposed MoE framework outperforms traditional static fusion methods and unimodal baselines in terms of prediction accuracy and adaptability, highlighting its potential in practical THz ISAC systems with ultra-massive multiple-input multiple-output (MIMO).

\end{abstract}

\begin{IEEEkeywords} 6G, beam prediction, deep learning, integrated sensing and communication, mixture of experts, THz communications.
\end{IEEEkeywords}

\section{Introduction}

Wireless communication systems operating at millimeter-wave (mmWave) and terahertz (THz) frequency bands have emerged as promising technologies to achieve ultra-high data rates required by future wireless applications, including 6G and beyond~\cite{letaief2019roadmap,letaief2021edge}.
Despite their enormous potential, mmWave and THz frequencies suffer from severe path loss and molecular absorption, necessitating massive and ultra-massive multiple-input multiple-output (MIMO) antenna arrays that establish highly directional beams to compensate for propagation losses, and provide the spatial resolution needed for sensing-communication synergy \cite{yu2025deep}.
However, the highly directional nature of THz beams poses substantial beam management challenges, particularly in dynamic environments with high mobility, such as vehicle-to-infrastructure (V2I) networks, drone-based communication systems, and augmented reality (AR) applications. In these cases, optimal beam directions must be frequently updated, resulting in substantial beam training overhead and latency, which become critical bottlenecks limiting the practical deployment of mobile THz communication systems \cite{yu2024ai}.

Several recent works have focused on reducing beam training overhead by employing classical methods, such as adaptive beam codebooks~\cite{noh2017multi}, sparsity-based compressive channel estimation~\cite{li2017millimeter}, and beam tracking algorithms~\cite{lim2021deep}. Nevertheless, these methods suffer from high beam training overhead that scales unfavorably with increasing antenna array sizes and user mobility, thereby limiting their applicability in dynamic THz communication scenarios.
Recently, deep learning-based approaches have been explored to proactively predict beam directions by leveraging environmental context information, including user positions~\cite{morais2023position}, camera images~\cite{alrabeiah2020millimeter}, radar signatures~\cite{demirhan2022radar}, and lidar measurements~\cite{jiang2022lidar}.
However, relying on a single sensing modality often results in suboptimal beam prediction performance, as each modality has its inherent limitations.
For example, camera images are sensitive to lighting and weather variations; radar and lidar measurements suffer from noise and clutter; and positional data collected by GPS typically lack sufficient accuracy for precise THz beam alignment.

To overcome these challenges, multimodal sensing has emerged as a promising approach, which integrates complementary information from multiple sensors (e.g., vision, radar, lidar, and positioning) to enhance the accuracy and robustness of beam prediction \cite{shi2024multimodal,tian2023multimodal}.
Nevertheless, existing multimodal fusion methods typically adopt static or heuristic architectures, such as direct concatenation or simple averaging of features extracted from different sensing modalities \cite{zhu2025advancing,charan2022vision}.
These methods cannot adaptively weight each modality’s importance according to its quality and reliability, which limits their robustness and accuracy in real-world dynamic environments.

To address these issues, in this paper, we propose a novel multimodal mixture-of-experts (MoE) deep learning framework for proactive beam prediction in THz integrated sensing and communication (ISAC) systems.
Specifically, the proposed MoE architecture comprises multiple modality-specific expert networks, each tailored to extract discriminative features from individual sensing modalities (e.g., vision, radar, lidar, and positioning). These extracted features are then dynamically combined through an adaptive gating network, which learns to assign fusion weights based on each modality’s real-time reliability and relevance. In particular, the gating network evaluates instantaneous modality conditions, such as environmental variations and sensor uncertainties, enabling the MoE framework to adaptively prioritize the most reliable modalities for accurate beam prediction. Extensive simulations performed on real-world vehicle-to-infrastructure (V2I) datasets demonstrate that the proposed multimodal MoE approach outperforms conventional static fusion methods and single-modality baselines in terms of prediction accuracy and robustness to environmental changes.

\emph{Notations:} Column vectors and matrices are denoted by boldface lowercase and boldface capital letters, respectively.
The symbol $\b{R}$ denotes the set of real numbers.
$\b{C}^{M\times N}$ represents the space of $M \times N$ complex-valued matrices.
$(\cdot)^\s{T}$ and $(\cdot)^\s{H}$ stand for the transpose and the conjugate transpose of their arguments, respectively.
$\b{E}[\cdot]$ denotes the expectation operation. $\nabla$ represents the gradient operator. $|\cdot|$ and $\|\cdot\|$ stand for the $\ell_1$ and $\ell_2$ norm of vectors, respectively.

\begin{figure}[t]
	\renewcommand\figurename{Fig.}
	\centering \vspace*{3pt} \setlength{\baselineskip}{10pt}
	\includegraphics[width = 0.49\textwidth,trim=5 0 0 0,clip]{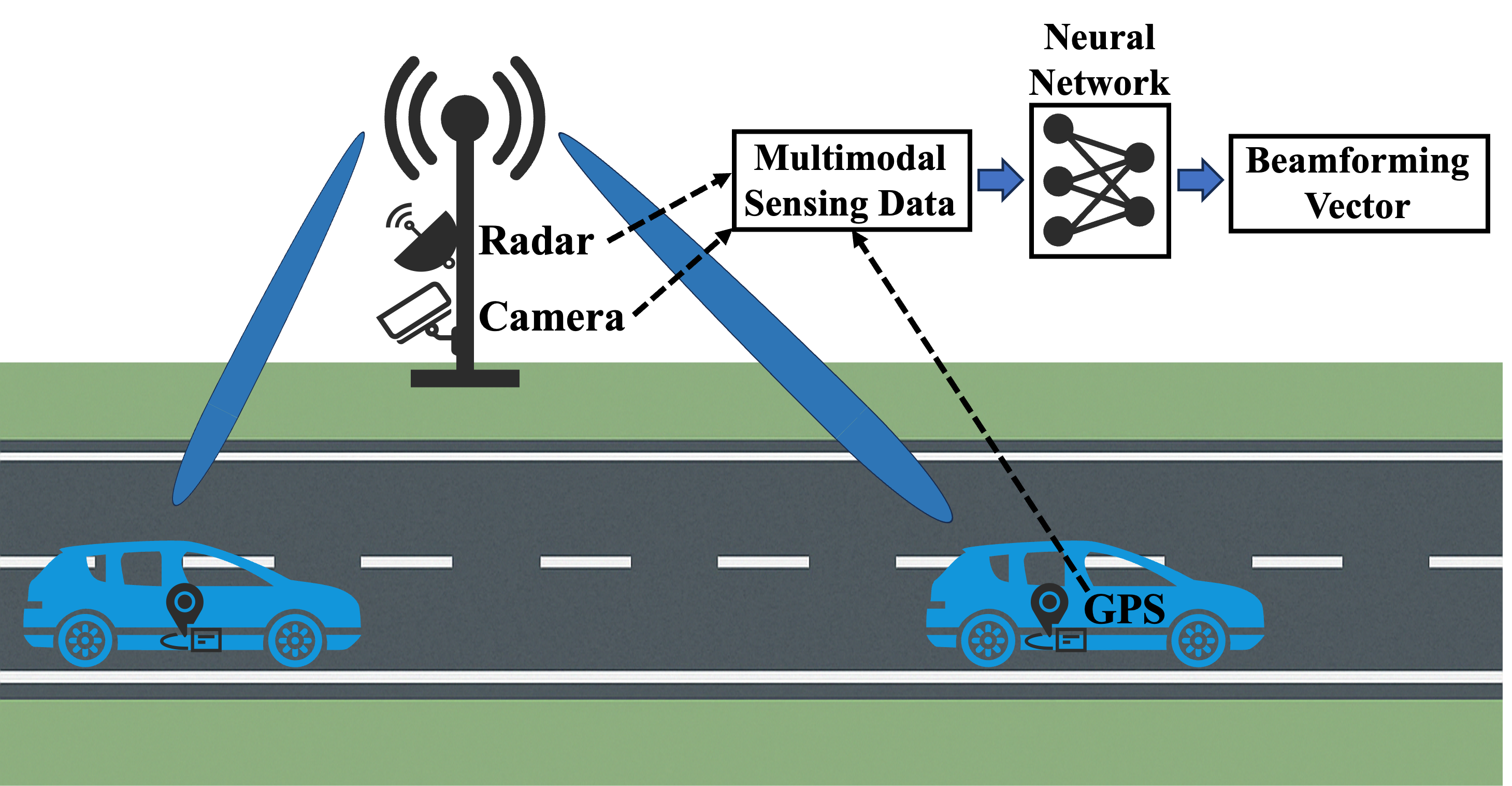}
	\caption{Illustration of multimodal sensing data empowered V2I ISAC system.}\label{fig_illus}
\end{figure}

\section{System Model and Problem Formulation}
\subsection{System Model}

We consider a downlink THz V2I communication scenario in which a roadside base station (BS), equipped with an $N$-element antenna array and multimodal sensors, serves a single-antenna mobile vehicle, as shown in Fig.\,\ref{fig_illus}.
Due to the significant path loss at THz frequencies, the BS employs directional beamforming to enhance the received signal strength and coverage range.
Specifically, the BS adopts a predefined beamforming codebook $\mathcal{F} = \{\mathbf{f}_m\}_{m=1}^{M}$, where $\mathbf{f}_m \in \mathbb{C}^{N\times 1}$ denotes the beamforming vector and $M$ represents the number of candidate beams. At each discrete time slot $t$, if the BS selects beamforming vector $\mathbf{f}(t)\in \mathcal{F}$, the downlink received signal at the user is given by
\begin{equation}\label{eq:received_signal}
	y(t)=\mathbf{h}(t)^\mathrm{H}\mathbf{f}(t)s(t)+z(t),
\end{equation}
where $\mathbf{h}(t)\in\mathbb{C}^{N\times 1}$ represents the instantaneous THz channel vector, $s(t)\in\mathbb{C}$ is the transmitted complex symbol satisfying $\mathbb{E}[|s(t)|^2] = 1$, and $z(t)\sim\mathcal{CN}(0,\sigma_0^2)$ denotes the additive white Gaussian noise.

The critical challenge in THz ISAC systems is selecting the optimal beamforming vector $\mathbf{f}^{\star}(t)\in\mathcal{F}$ at each time slot $t$, due to the narrow beamwidth at THz frequencies.
The optimal beam $\mathbf{f}^\star(t)$ at time slot $t$ is identified by maximizing the effective received power (beamforming gain), which can be formulated as 
\begin{equation}
	\begin{aligned}
		\mathbf{f}^\star(t)=\argmax_{\mathbf{f}\in\mathcal{F}} |\mathbf{h}(t)^\mathrm{H}\mathbf{f}|^2.
	\end{aligned}
\end{equation}
Conventionally, optimal beam training requires an exhaustive search of all candidate beams with excessive communication overhead and latency, which is particularly detrimental in high-mobility THz systems that require frequent beam alignment.
To alleviate this issue, the BS can leverage synchronized multimodal sensing data collected from multiple sensors for proactive beam prediction, such as RGB images from cameras, positional data from GPS, radar, and lidar measurements. Specifically, at each time slot $t$, the collected multimodal sensing data are represented as:
\begin{equation}
	\mathbf{X}(t)=\{\mathbf{X}_d(t)\}_{d=1}^D,
\end{equation}
where each $\mathbf{X}_d(t)$ corresponds to the measurements from a specific sensing modality. By integrating the multimodal sensory information, the BS proactively estimates the optimal beamforming vector without the need for explicit beam training, thereby reducing the communication overhead and latency associated with conventional beam alignment methods.

\begin{figure*}[t]
	\renewcommand\figurename{Fig.}
	\centering \vspace*{1pt} \setlength{\baselineskip}{10pt}
	\includegraphics[width = 0.99\textwidth,trim=0 0 0 0,clip]{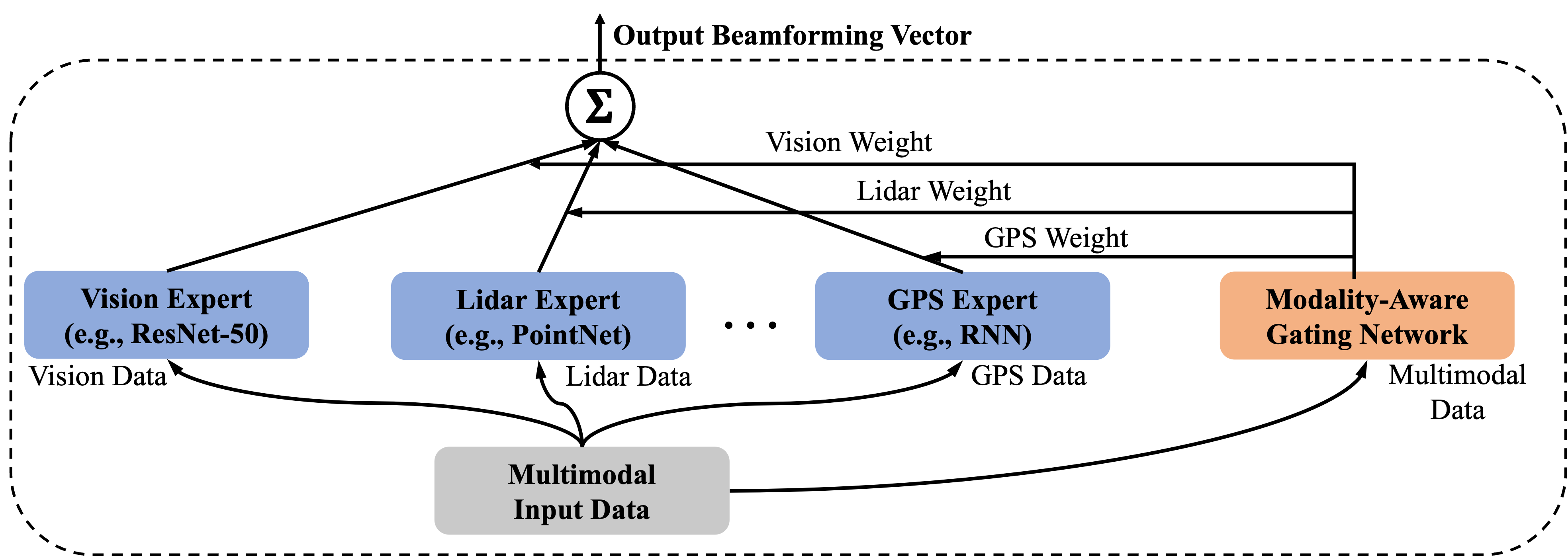}
	\caption{The proposed multimodal MoE framework for proactive beam prediction, consisting of modality-specific experts and a gating network to dynamically fuse multimodal features.}\label{fig_moe}
\end{figure*}

\subsection{Problem Formulation}

Given the multimodal sensing-aided THz V2I communication system described above, our goal is to proactively predict the optimal beamforming vector at each time instance without incurring beam training overhead. Specifically, we aim to design a predictive deep neural network, parameterized by $\boldsymbol{\Theta}$, which utilizes multimodal sensing data to estimate the optimal beamforming direction from a predefined beam codebook.
Formally, the multimodal sensing-aided beam prediction problem can be formulated as follows,
\begin{equation}
	\begin{aligned}
		\max_{\boldsymbol{\Theta}} &~~ \mathbb{E}{}\left[  |\mathbf{h}(t)^\mathrm{H}\mathbf{\hat f}(t)|^2 \right] \\
		\textup{s.t.}\,& ~~ \hat{\mathbf{f}}(t)=g_{\boldsymbol{\Theta}}(\mathbf{X}(t)),
	\end{aligned}
\end{equation}
where $\hat{\mathbf{f}}(t)$ is the beamforming vector selected by the predictive model at time slot $t$, the expectation $\mathbb{E}[\cdot]$ captures the statistical nature of multimodal sensing data and channel variations, and $g_{\boldsymbol{\Theta}}(\cdot)$ represents the learning model that maps multimodal sensing data $\mathbf{X}(t)$ to the predicted beamforming vector.
Deep learning-based multimodal sensing allows the BS to proactively select the optimal beamforming vector that maximizes the received signal strength. This can significantly reduce the latency and overhead of beam training and enhance the robustness in high-mobility THz systems.

To address the multimodal beam prediction problem, existing methods generally fall into two categories: end-to-end neural fusion methods and heuristic feature fusion methods.
End-to-end neural fusion approaches employ a single, unified neural network architecture that directly maps synchronized multimodal sensory data to the beam prediction output. Although these methods benefit from simplicity and strong representation capabilities, they inherently perform fusion in a black-box manner, failing to account for the varying reliability and quality of individual sensing modalities. In contrast, heuristic feature fusion methods explicitly incorporate modality-specific reliability through predefined heuristic strategies, such as weighted summation or direct concatenation of features independently extracted from each sensing modality. While offering better interpretability, these heuristic methods depend heavily on manually defined fusion criteria, resulting in limited adaptability and suboptimal performance in dynamic environments.

Motivated by these limitations, we propose a novel multimodal MoE deep learning framework, which dynamically combines modality-specific predictions through an adaptive gating mechanism. As detailed in the next section, the proposed MoE approach explicitly models the varying reliability of each sensing modality, enabling more accurate and robust beam predictions in dynamic THz ISAC environments.

\section{Proposed Multimodal MoE Framework}

In this section, we propose a multimodal MoE deep learning framework for proactive beam prediction in THz ISAC systems. We first introduce the multimodal MoE architecture, consisting of modality-specific expert networks and a modality-aware gating network. Then, we present the complete training procedure of the proposed MoE framework.

\subsection{Multimodal Mixture-of-Experts Architecture}
The multimodal MoE framework consists of multiple expert networks, each explicitly dedicated to extracting features from a specific sensing modality, as shown in Fig.\,\ref{fig_moe}. Each modality-specific expert is implemented as a deep neural network to effectively capture intrinsic characteristics unique to that modality, generating discriminative high-level representations for subsequent fusion.
Specifically, given multimodal sensory input $\mathbf{X}(t)=\{\mathbf{X}_d(t)\}_{d=1}^D$ at time slot $t$, we construct a set of expert networks $\{f_d(\cdot;\boldsymbol{\theta}_d)\}_{d=1}^D$ where $\boldsymbol{\theta}_d$ denotes the learnable parameters of the expert for the $d$-th modality. Each expert network receives input from one sensing modality and generates modality-specific feature embeddings, which can be formally expressed as
\begin{equation}
	\mathbf{z}_{d}(t)=f_d(\mathbf{X}_d(t);\boldsymbol{\theta}_d), \forall d.
\end{equation}
As illustrative examples, the radar expert can employ convolutional or recurrent neural networks to effectively capture attributes such as range, angle, and velocity from radar measurements. The lidar expert typically utilizes point-cloud-oriented architectures (e.g., PointNet \cite{qi2017pointnet} or PointNet++ \cite{qi2017pointnet++}) to extract geometric features from lidar data. Furthermore, the visual expert may adopt established CNN architectures (e.g., ResNet \cite{he2016deep} or Vision Transformer \cite{dosovitskiy2020vit}) to encode semantically rich contextual information from RGB images. By employing modality-specific neural architectures, the proposed MoE framework effectively captures intrinsic modality features and enhances the robustness and discriminative capability of the fused multimodal representation.

However, extracting features independently from each modality neglects the inherent interactions and complementary relationships among different sensing modalities.
In practice, the reliability and quality of each modality varies dynamically due to environmental conditions, sensor limitations, and operating environments.
Thus, statically or heuristically fusing modality-specific representations might degrade overall predictive accuracy and robustness.
To overcome this limitation, we introduce a modality-aware gating network that dynamically generates fusion weights by explicitly assessing the instantaneous reliability of each sensing modality. In the following subsection, we elaborate on the design of this gating network and highlight its role in enabling adaptive and interpretable multimodal feature fusion.

\subsection{Modality-Aware Gating Network}
The modality-aware gating network serves as a critical component of the proposed multimodal MoE framework. Since the reliability and contribution of each sensing modality may vary under dynamic environmental conditions, assigning fixed or equal weights to modality-specific representations can degrade predictive performance. To overcome this limitation, we propose a gating network that adaptively evaluate the instantaneous reliability and relevance of each modality and generate corresponding fusion weights.

Specifically, the gating network is represented by a learnable function $f_g(\cdot;\boldsymbol{\theta}_g)$, parameterized by $\boldsymbol{\theta}_g$, which takes the synchronized multimodal input $\mathbf{X}(t)=\{\mathbf{X}_d(t)\}_{d=1}^D$ and outputs normalized fusion weights
\begin{equation}
	[w_{1}(t), w_{2}(t), \dots, w_{D}(t)] = \text{softmax}(f_g(\mathbf{X}(t);\boldsymbol{\theta}_g)),
\end{equation}
where the softmax activation ensures the non-negativity and normalization of weights, i.e.,
\begin{equation}
	\begin{aligned}
	\sum_{d=1}^{D} w_d(t) = 1, \\
	w_d(t)\geq 0, \forall d.
	\end{aligned}
\end{equation}
The gating network first aggregates multimodal sensing inputs into a compact intermediate representation, effectively capturing cross-modality interactions and dependencies. This intermediate representation is then mapped into modality-specific scores via trainable neural network layers with nonlinear activations (e.g., ReLU). Finally, these scores are transformed into normalized weights by the softmax operation, explicitly reflecting the instantaneous importance of each sensing modality. The gating network architecture (e.g., fully-connected, convolutional, or attention-based layers) can be selected based on the characteristics of modalities, data complexity, and empirical performance. For instance, convolutional or attention-based layers may be employed for visual modalities, enabling efficient extraction of rich semantic features.

\begin{algorithm}[!tbp]
	\caption{Training Procedure of the Proposed Multimodal MoE Framework.}
	\label{algo:multimodal_moe}
	\KwIn{Training dataset $\mathcal{D}=\{(\mathbf{X}(t),\mathbf{f}(t))\}_{t=1}^{T}$, initialized expert network parameters $\{\boldsymbol{\theta}_d\}_{d=1}^{D}$, gating network parameters $\boldsymbol{\theta}_g$, output prediction network parameters $\boldsymbol{\theta}_o$, learning rate $\eta$.}
	\KwOut{Optimized parameters $\{\boldsymbol{\theta}_d\}_{d=1}^{D}$, $\boldsymbol{\theta}_g$, and $\boldsymbol{\theta}_o$.}
	
	Randomly initialize expert parameters $\{\boldsymbol{\theta}_d\}_{d=1}^{D}$, gating parameters $\boldsymbol{\theta}_g$, and output parameters $\boldsymbol{\theta}_o$\;
	
	\For{epoch $=1,2,\dots, E$}{
		Shuffle the training dataset $\mathcal{D}$\;
		\For{each training sample $(\mathbf{X}(t),\mathbf{f}(t))\in\mathcal{D}$}{
			Compute modality-specific expert embeddings:
			\vspace{-4pt}
			\begin{equation}
				\vspace{-2pt}
				\begin{aligned}
					\mathbf{z}_{d}(t)=f_d(\mathbf{X}_d(t);\boldsymbol{\theta}_d),~\forall d;
				\end{aligned}
			\end{equation}
			
			Compute modality-aware fusion weights:
			\vspace{-3pt}
			\begin{equation}
				\begin{aligned}
					[w_{1}(t),\dots,w_{D}(t)]=\text{softmax}(f_g(\mathbf{X}(t);\boldsymbol{\theta}_g));
				\end{aligned}
			\end{equation}
			
			Compute fused multimodal representation:
			\vspace{-6pt}
			\begin{equation}
				\vspace{-4pt}
				\begin{aligned}
					\mathbf{z}(t)=\sum_{d=1}^{D} w_d(t)\mathbf{z}_{d}(t);
				\end{aligned}
			\end{equation}
			
			Compute final prediction output:
			\vspace{-3pt}
			\begin{equation}
				\vspace{-2pt}
				\begin{aligned}
					\hat{\mathbf{f}}(t)=f_o(\mathbf{z}(t);\boldsymbol{\theta}_o);
				\end{aligned}
			\end{equation}
			
			Evaluate supervised loss function:
			\vspace{-4pt}
			\begin{equation}
				\vspace{-3pt}
				\begin{aligned}
					\mathcal{L}_t=\mathcal{L}(\mathbf{f}(t),\hat{\mathbf{f}}(t));
				\end{aligned}
			\end{equation}
			
			Compute gradients via backpropagation:
			\vspace{-4pt}
			\begin{equation}
				\vspace{-3pt}
				\begin{aligned}
					\nabla_{\boldsymbol{\theta}_d}\mathcal{L}_t,\forall d;\quad\nabla_{\boldsymbol{\theta}_g}\mathcal{L}_t;\quad\nabla_{\boldsymbol{\theta}_o}\mathcal{L}_t;
				\end{aligned}
			\end{equation}
			
			Update parameters with gradient descent:
			\vspace{-4pt}
			\begin{equation}
				\vspace{-3pt}
				\begin{aligned}
					\boldsymbol{\theta}_d&\leftarrow\boldsymbol{\theta}_d-\eta\nabla_{\boldsymbol{\theta}_d}\mathcal{L}_t,~\forall d ;\\
					\boldsymbol{\theta}_g&\leftarrow\boldsymbol{\theta}_g-\eta\nabla_{\boldsymbol{\theta}_g}\mathcal{L}_t;\\
					\boldsymbol{\theta}_o&\leftarrow\boldsymbol{\theta}_o-\eta\nabla_{\boldsymbol{\theta}_o}\mathcal{L}_t;
				\end{aligned}
			\end{equation}
		}
	}
\end{algorithm}

The final fused multimodal representation $\mathbf{z}(t)$ at time slot $t$ is computed as a weighted combination of the modality-specific expert outputs $\{\mathbf{z}_d(t)\}_{d=1}^D$, given by
\begin{equation}
	\mathbf{z}(t) = \sum_{d=1}^{D} w_d(t)\mathbf{z}_{d}(t).
\end{equation}
This fused representation $\mathbf{z}(t)$ is subsequently input to a dedicated beam prediction network, which directly maps it to the predicted optimal beamforming vector selected from the predefined beam codebook $\mathcal{F}$. Specifically, the predicted beamforming vector $\hat{\mathbf{f}}(t)$ is obtained via
\begin{equation}
	\hat{\mathbf{f}}(t) = f_{o}(\mathbf{z}(t);\boldsymbol{\theta}_{o}),
\end{equation}
where $f_{o}(\cdot;\boldsymbol{\theta}_{o})$ denotes the beam prediction network parameterized by learnable parameters $\boldsymbol{\theta}_{o}$.

Through joint end-to-end training of the modality-specific expert networks $\{f_d(\cdot;\boldsymbol{\theta}_d)\}_{d=1}^D$, the modality-aware gating network $f_g(\cdot;\boldsymbol{\theta}_g)$, and the beam prediction network $f_{o}(\cdot;\boldsymbol{\theta}_{o})$, the proposed multimodal MoE framework dynamically adapts the fusion weights according to instantaneous modality reliability. This adaptive mechanism allows the model to effectively exploit complementary multimodal information, significantly enhancing beam prediction accuracy and robustness in dynamic THz ISAC environments.

\subsection{Algorithm Development}

In this subsection, we present the detailed training procedure of the proposed multimodal MoE framework. Specifically, our goal is to jointly optimize the parameters of the modality-specific expert networks, the modality-aware gating network, and the subsequent beam prediction network. To achieve this, we formulate the training as a supervised learning problem using a labeled multimodal dataset.
Given a training dataset consisting of synchronized multimodal sensing inputs and corresponding ground-truth labels, denoted as $\mathcal{D}=\{(\mathbf{X}(t), \mathbf{f}(t))\}_{t=1}^{T}$, we define a supervised loss function $\mathcal{L}(\cdot)$ to measure the discrepancy between the predicted beamforming vector $\hat{\mathbf{f}}(t)$ and the ground-truth label $\mathbf{f}(t)$.
Formally, the training objective is defined as
\begin{equation}
	\min_{\{\boldsymbol{\theta}_d\}_{d=1}^D,\boldsymbol{\theta}_g,\boldsymbol{\theta}_o}~\frac{1}{T}\sum_{t=1}^{T}\mathcal{L}\bigl(\mathbf{f}(t), \hat{\mathbf{f}}(t)\bigr),
\end{equation}
where the prediction $\hat{\mathbf{f}}(t)$ is obtained by sequentially computing expert outputs, modality-aware fusion weights, fused representations, and final predictions as
\begin{equation}
	\begin{aligned}
		\mathbf{z}_{d}(t)&=f_d(\mathbf{X}_d(t);\boldsymbol{\theta}_d), ~\forall d,\\[3pt]
		[w_{1}(t),\dots,w_{D}(t)]&=\text{softmax}(f_g(\mathbf{X}(t);\boldsymbol{\theta}_g)),\\[3pt]
		\mathbf{z}(t)&=\sum_{d=1}^{D}w_d(t)\mathbf{z}_d(t),\\[3pt]
		\hat{\mathbf{f}}(t)&=f_o(\mathbf{z}(t);\boldsymbol{\theta}_o).
	\end{aligned}
\end{equation}

The complete training algorithm is summarized in {Algorithm~\ref{algo:multimodal_moe}}. By jointly optimizing the expert networks, gating network, and beam prediction network, the proposed multimodal MoE framework collaboratively learns to capture modality-specific contributions and interactions, thus improving predictive accuracy and robustness across diverse multimodal THz ISAC scenarios.

\section{Simulation Results}

\subsection{Simulation Setups}

In this section, we evaluate the proposed multimodal MoE framework by using two real-world V2I ISAC scenarios (Scenario 2 and Scenario 8) from the publicly available DeepSense6G dataset \cite{alkhateeb2023deepsense}. The testbed in these scenarios comprises a single-antenna mobile vehicle and a stationary BS equipped with a 16-element phased array antenna with an oversampled codebook of 64 predefined beams.
The BS is equipped with an RGB-D camera capturing RGB images at 960×540 resolution, while the mobile vehicle unit is equipped with a GPS-RTK receiver providing real-time positional data (i.e., latitude and longitude). Specifically, Scenario 2 captures nighttime conditions, while Scenario 8 represents daytime conditions. In the simulations, we leverage synchronized RGB images from the BS and GPS data from the vehicle to proactively predict the optimal beam direction from the predefined beam codebook.

In the proposed multimodal MoE architecture, the vision expert is implemented using a ResNet-18 network \cite{he2016deep}, and the GPS expert employs a two-layer multilayer perceptron (MLP). The modality-aware gating network is designed as a lightweight three-layer MLP to dynamically generate fusion weights based on the instantaneous reliability of each modality.
We compare the proposed multimodal MoE framework with the following three benchmark methods:

\begin{itemize}
	\item \textbf{Vision-Only:} This method utilizes only the RGB camera images captured at the stationary BS to predict the optimal beam.
	\item \textbf{Position-Only:} This approach leverages only the vehicle’s GPS location data to determine the optimal beam direction.
	\item \textbf{Feature Concatenation Fusion:} This baseline independently extracts features from both the RGB images and GPS data, concatenates these features into a unified representation, and inputs it into a single predictive network for beam prediction.
\end{itemize}

\begin{figure}[t]
	\renewcommand\figurename{Fig.}
	\centering \vspace*{1pt} \setlength{\baselineskip}{10pt}
	\includegraphics[width = 0.49\textwidth,trim=10 0 0 0,clip]{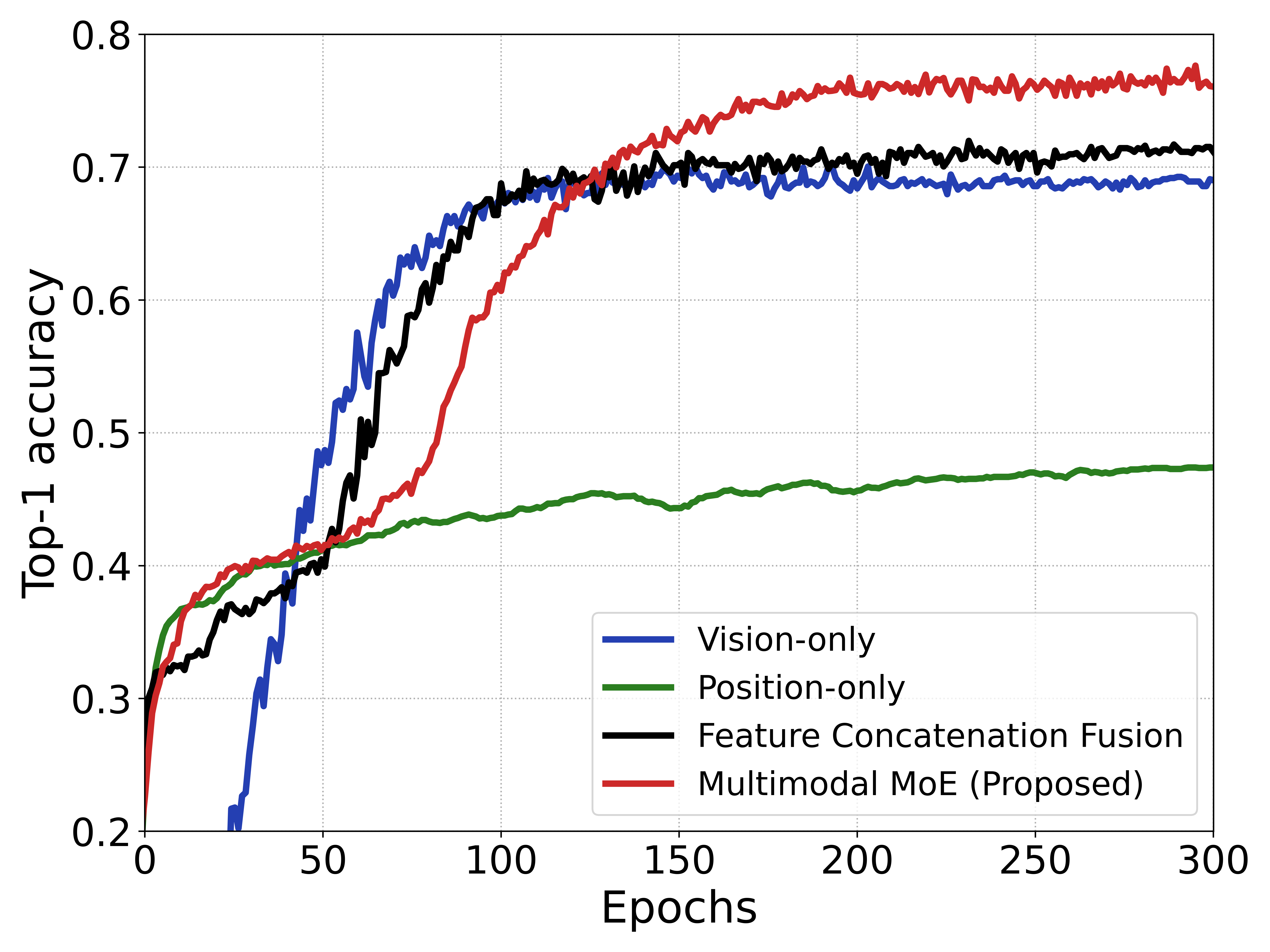}
	\caption{Top-1 accuracy comparison of the proposed multimodal MoE framework against benchmark methods.}
\end{figure}

\begin{figure*}[t]
	\renewcommand\figurename{Fig.}
	\centering \vspace*{1pt} \setlength{\baselineskip}{10pt}
	\includegraphics[width = 0.99\textwidth,trim=10 10 10 0,clip]{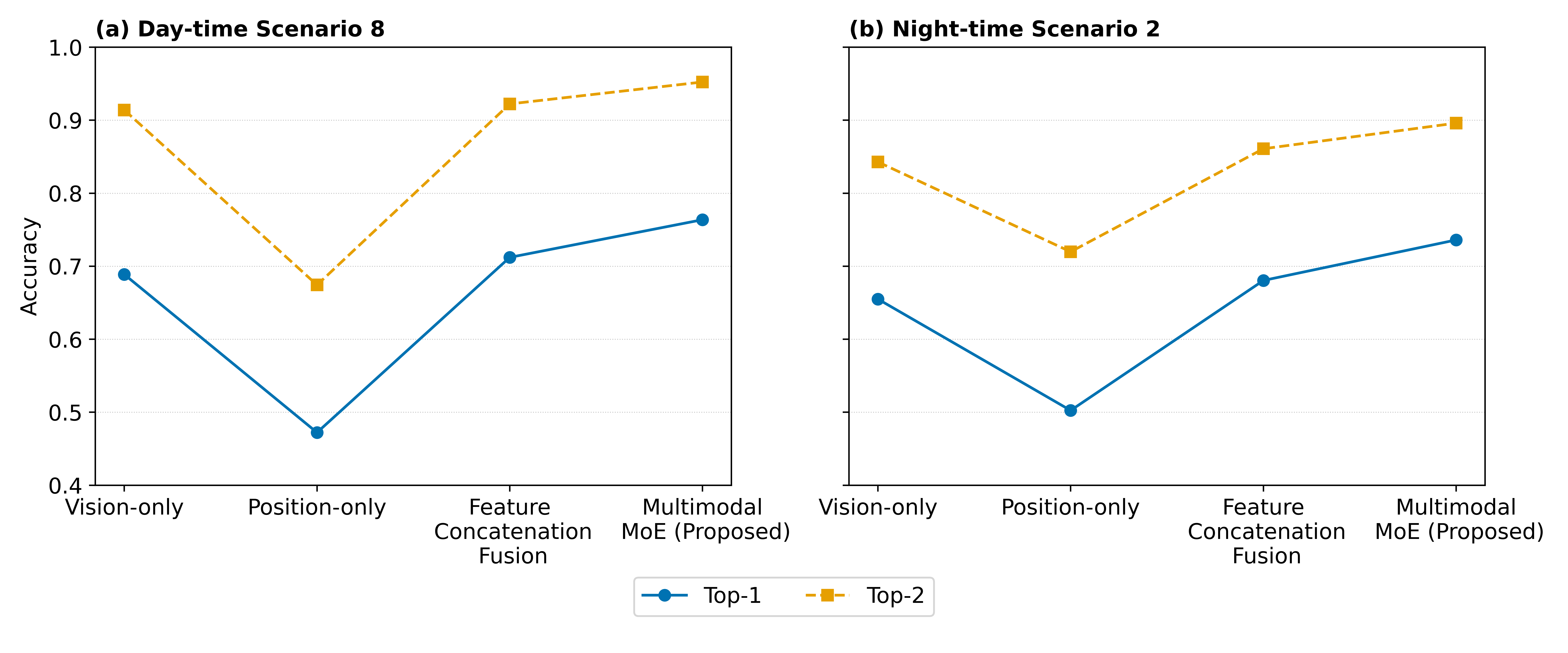}
	\vspace*{-6pt}
	\caption{Comparison of top-1 and top-2 beam prediction accuracy in (a) Scenario 8 (daytime) and (b) Scenario 2 (nighttime) among the proposed multimodal MoE framework and baseline approaches.}
\end{figure*}



\subsection{Performance Evaluation}

Fig.\,3 demonstrates the top-1 beam prediction accuracy achieved by different approaches on Scenario 8 in the DeepSense6G dataset. It is evident that the proposed multimodal MoE method outperforms all baseline methods after approximately 100 epochs, achieving higher prediction accuracy with improved training stability. In contrast, the unimodal methods exhibit limited accuracy due to their inability to fully exploit complementary multimodal features, while the feature concatenation fusion method provides moderate improvements but remains inferior to the adaptive fusion capability of the proposed multimodal MoE approach.

In Fig.\,4, we further compare the top-1 and top-2 beam prediction accuracy of all methods in both Scenario 8 (daytime) and Scenario 2 (nighttime). The position-only method consistently yields the lowest performance due to the inherent inaccuracies of positional information for precise THz beam alignment. The vision-only method achieves relatively good performance during the daytime but experiences notable performance degradation at night, indicating its sensitivity to environmental lighting conditions. The feature concatenation fusion approach partially addresses the limitations of single-modality methods by combining complementary features, thereby improving overall accuracy. Nevertheless, the proposed multimodal MoE framework consistently achieves the highest top-1 and top-2 accuracy across both scenarios, clearly demonstrating its robustness and superior capability of adaptively integrating multimodal information under varying environmental conditions.

\section{Conclusion}

In this paper, we investigated multimodal sensing-aided beam prediction for THz ISAC systems. To overcome the limitations of conventional beam training and static multimodal fusion methods, we proposed a novel multimodal MoE deep learning framework.
The proposed MoE framework employed modality-specific expert networks to extract complementary features, and dynamically fused them using adaptive weights generated by a gating network according to instantaneous modality reliability.
Simulation results on real-world V2I dataset demonstrated that the proposed MoE framework significantly outperformed static fusion methods and unimodal baselines in terms of prediction accuracy and adaptability, highlighting its potential for adaptive beam prediction in practical THz ISAC MIMO systems.


\bibliographystyle{ieeetr}
\bibliography{IEEEabrv,refs}

\begin{thebibliography}{10}

\bibitem{letaief2019roadmap}
K.~B. Letaief, W.~Chen, Y.~Shi, J.~Zhang, and Y.-J.~A. Zhang, ``The roadmap to
  {6G}: {AI} empowered wireless networks,'' {\em IEEE Commun. Mag.}, vol.~57,
  no.~8, pp.~84--90, 2019.

\bibitem{letaief2021edge}
K.~B. Letaief, Y.~Shi, J.~Lu, and J.~Lu, ``Edge artificial intelligence for
  {6G}: Vision, enabling technologies, and applications,'' {\em IEEE J. Sel.
  Areas Commun.}, vol.~40, no.~1, pp.~5--36, 2021.

\bibitem{yu2025deep}
W.~Yu, Y.~Ma, H.~He, S.~Song, J.~Zhang, and K.~B. Letaief, ``Deep learning for
  near-field {XL-MIMO} transceiver design: Principles and techniques,'' {\em
  IEEE Commun. Mag.}, vol.~63, no.~1, pp.~52--58, 2025.

\bibitem{yu2024ai}
W.~Yu, H.~He, S.~Song, J.~Zhang, L.~Dai, L.~Zheng, and K.~B. Letaief, ``{AI}
  and deep learning for {THz} ultra-massive {MIMO}: From model-driven
  approaches to foundation models,'' {\em arXiv preprint arXiv:2412.09839},
  2024.

\bibitem{noh2017multi}
S.~Noh, M.~D. Zoltowski, and D.~J. Love, ``Multi-resolution codebook and
  adaptive beamforming sequence design for millimeter wave beam alignment,''
  {\em IEEE Trans. Wireless Commun.}, vol.~16, no.~9, pp.~5689--5701, 2017.

\bibitem{li2017millimeter}
X.~Li, J.~Fang, H.~Li, and P.~Wang, ``Millimeter wave channel estimation via
  exploiting joint sparse and low-rank structures,'' {\em IEEE Trans. Wireless
  Commun.}, vol.~17, no.~2, pp.~1123--1133, 2017.

\bibitem{lim2021deep}
S.~H. Lim, S.~Kim, B.~Shim, and J.~W. Choi, ``Deep learning-based beam tracking
  for millimeter-wave communications under mobility,'' {\em IEEE Trans.
  Commun.}, vol.~69, no.~11, pp.~7458--7469, 2021.

\bibitem{morais2023position}
J.~Morais, A.~Bchboodi, H.~Pezeshki, and A.~Alkhateeb, ``Position-aided beam
  prediction in the real world: How useful {GPS} locations actually are?,'' in
  {\em ICC 2023-IEEE Int. Conf. Commun.}, pp.~1824--1829, IEEE, 2023.

\bibitem{alrabeiah2020millimeter}
M.~Alrabeiah, A.~Hredzak, and A.~Alkhateeb, ``Millimeter wave base stations
  with cameras: Vision-aided beam and blockage prediction,'' in {\em 2020 IEEE
  Veh. Technol. Conf. (VTC2020-Spring)}, pp.~1--5, IEEE, 2020.

\bibitem{demirhan2022radar}
U.~Demirhan and A.~Alkhateeb, ``Radar aided {6G} beam prediction: Deep learning
  algorithms and real-world demonstration,'' in {\em 2022 IEEE Wireless Commun.
  Netw. Conf. (WCNC)}, pp.~2655--2660, IEEE, 2022.

\bibitem{jiang2022lidar}
S.~Jiang, G.~Charan, and A.~Alkhateeb, ``Lidar aided future beam prediction in
  real-world millimeter wave {V2I} communications,'' {\em IEEE Wireless Commun.
  Lett.}, vol.~12, no.~2, pp.~212--216, 2022.

\bibitem{shi2024multimodal}
B.~Shi, M.~Li, M.-M. Zhao, M.~Lei, and L.~Li, ``Multimodal deep learning
  empowered millimeter-wave beam prediction,'' in {\em 2024 IEEE Veh. Technol.
  Conf. (VTC2024-Spring)}, pp.~1--6, IEEE, 2024.

\bibitem{tian2023multimodal}
Y.~Tian, Q.~Zhao, F.~Boukhalfa, K.~Wu, F.~Bader, {\em et~al.}, ``Multimodal
  transformers for wireless communications: A case study in beam prediction,''
  {\em arXiv preprint arXiv:2309.11811}, 2023.

\bibitem{zhu2025advancing}
Q.~Zhu, Y.~Wang, W.~Li, H.~Huang, and G.~Gui, ``Advancing multi-modal beam
  prediction with cross-modal feature enhancement and dynamic fusion
  mechanism,'' {\em IEEE Trans. Commun.}, 2025.

\bibitem{charan2022vision}
G.~Charan, T.~Osman, A.~Hredzak, N.~Thawdar, and A.~Alkhateeb,
  ``Vision-position multi-modal beam prediction using real millimeter wave
  datasets,'' in {\em 2022 IEEE Wireless Commun. Netw. Conf. (WCNC)},
  pp.~2727--2731, IEEE, 2022.

\bibitem{qi2017pointnet}
C.~R. Qi, H.~Su, K.~Mo, and L.~J. Guibas, ``{PointNet}: Deep learning on point
  sets for {3D} classification and segmentation,'' in {\em Proc. IEEE Conf.
  Comput. Vis. Pattern Recognit. (CVPR)}, pp.~652--660, 2017.

\bibitem{qi2017pointnet++}
C.~R. Qi, L.~Yi, H.~Su, and L.~J. Guibas, ``{PointNet++}: Deep hierarchical
  feature learning on point sets in a metric space,'' in {\em Proc. Adv. Neural
  Inf. Process. Syst. (NeurIPS)}, pp.~5099--5108, 2017.

\bibitem{he2016deep}
K.~He, X.~Zhang, S.~Ren, and J.~Sun, ``Deep residual learning for image
  recognition,'' in {\em Proc. IEEE Conf. Comput. Vis. Pattern Recognit.
  (CVPR)}, pp.~770--778, 2016.

\bibitem{dosovitskiy2020vit}
A.~Dosovitskiy, L.~Beyer, A.~Kolesnikov, D.~Weissenborn, X.~Zhai,
  T.~Unterthiner, M.~Dehghani, M.~Minderer, G.~Heigold, S.~Gelly, J.~Uszkoreit,
  and N.~Houlsby, ``An image is worth 16x16 words: Transformers for image
  recognition at scale,'' in {\em Proc. Int. Conf. Learn. Representations
  (ICLR)}, 2021.

\bibitem{alkhateeb2023deepsense}
A.~Alkhateeb, G.~Charan, T.~Osman, A.~Hredzak, J.~Morais, U.~Demirhan, and
  N.~Srinivas, ``Deepsense {6G}: A large-scale real-world multi-modal sensing
  and communication dataset,'' {\em IEEE Commun. Mag.}, vol.~61, no.~9,
  pp.~122--128, 2023.

\end{thebibliography}


\end{document}